\documentclass[aps,prstab,twocolumn,groupedaddress]{revtex4-1}

\usepackage{graphicx}
\usepackage{subfigure}
\usepackage{comment}
\usepackage{verbatim}
\usepackage{url}

\usepackage{natbib}
\begin{document}

%\preprint{}

\title{Multi-objective Optimizations of a Novel Cryo-cooled DC Gun Based Ultra Fast Electron Diffraction Beamline}


\author{Colwyn Gulliford}\email{cg248@cornell.edu}
\author{Adam Bartnik}%\email{acb20@cornell.edu}
%\author{Jared Maxson}
\author{Ivan Bazarov}%\email{ib38@cornell.edu}
%\author{John Dobbins} 
%\author{Bruce Dunham}
%\author{Luca Cultrera}
%\author{Yulin Li}
%\author{Xianghong Liu} 
%\author{Karl Smolenski}
%\email[Colwyn Gulliford]{cg248}
%\thanks{}
\affiliation{CLASSE, Cornell University, 161 Synchrotron Drive
Ithaca, NY 14853-8001}


\date{\today}

\begin{abstract}
We present the results of multi-objective genetic algorithm optimizations of a potential single shot ultra fast electron diffraction beamline utilizing a 225 kV dc gun with a novel cryocooled photocathode system and buncher cavity.  Optimizations of the transverse projected emittance as a function of bunch charge are presented and discussed in terms of the scaling laws derived in the charge saturation limit.  Additionally, optimization of the transverse coherence length as a function of final rms bunch length at sample location have been performed for three different sample radii: $50$, $100$, and $200$ $\mu$m, for two final bunch charges: $10^5$ and $10^6$ electrons.  Analysis of the solutions is discussed, as are the effects of disorder induced heating.  In particular, a relative coherence length of $L_{c,x}/\sigma_x=$ 0.27 was obtained for a final bunch charge of $10^5$ electrons and final bunch length of $\sigma_t\approx 100$ fs.  For a final charge of $10^6$ electrons the cryogun produces  $L_{c,x}/\sigma_x\approx0.1$ nm/$\mu$m for $\sigma_t\approx 100-200$ fs and $\sigma_x\geq50$ $\mu$m.  These results demonstrate the viability of using genetic algorithms in the design and operation of ultrafast electron diffraction beamlines.

\end{abstract}

\pacs{PACS numbers?}
%\keywords{}


\maketitle

\section{Introduction}
  
The desire for single-shot ultrafast electron diffraction (UED) beamlines ($\sigma_t \lesssim$ 100 fs, $q\sim 10^6$ electrons) capable of imaging molecular and atomic motion continues to push the development of both photocathode and cold atom electron sources \cite{ref:uedUT1,ref:ued:dcbun2solconcept,ref:uedUT2,ref:ued:dcbun2sol2,ref:ued:dcbun2sol,ref:uedUT3,ref:coldatoms1,ref:coldatoms2}.  In the case of photoemission sources, advances in the development of low mean transverse energy (MTE) photocathodes \cite{ref:redmte,ref:coldcathode}, as well as both DC gun and normal conducting rf gun technology \cite{ref:pietro0}, now bring the goal of creating single shot electron diffraction beamlines with lengths on the order of meters with in reach.

For such devices, the required charge and beam sizes at the cathode imply transporting a strongly space charged dominated beam.  Building on the successful application of Multi-Objective Genetic Algorithm (MOGA) optimized simulations of space charge dominated beams used in the design and operation of the Cornell photoinjector \cite{ref:lowemitter,ref:lowemitter2,ref:lowemitter3}, we apply the same techniques to a moderate energy DC gun followed by two solenoids and a NCRF buncher cavity \cite{ref:ued:dcbun2solconcept,ref:ued:dcbun2sol,ref:ued:dcbun2sol2}.  We use the smallest MTEs considered achievable given the excellent vacuum environment provided by this gun technology.  In particular, recent work points to the ability to reduce the cathode MTE via cooling of the cathode \cite{ref:coldcathode}, and data suggests cathode MTEs as low as 5 meV (cathode emittance of 0.1 $\mu$m/mm) may be possible using multi-akali antimonide cathodes cooled to 20 K.  

This work is structured as follows: first, we briefly review the definition of coherence and the expected scaling with critical initial laser and beam parameters.  Next, a detailed description of the beamline set-up, and the parameters for optimization is given.  The results of an initial round of optimizations of the emittance vs. bunch charge, as well as detailed optimizations of the coherence length vs. final bunch length at several final beam sizes ($\sigma_x \approx $ 25, 50, 100 $\mu$m) and bunch charges ($10^5$ and $10^6$ electrons) follow.  From the optimal fronts, examples for $\sigma_x\approx 50$ $\mu$m are simualted for both final charges, and the dynamics in each case discussed.  

\subsection{Coherence Length From Photocathode Sources}

The  transverse coherence length is defined as $L_{c,x}  \approx \hbar/\sigma_{p_{x}}  =  \lambdabar_e/\sigma_{\gamma\beta_{x}}$ \cite{ref:uedUT1,ref:uedUT2,ref:uedUT3,ref:ued:dcbun2sol,ref:ued:dcbun2solconcept,ref:ued:dcbun2sol2,ref:uedmiller,ref:coldatoms1,ref:coldatoms2}
%\begin{eqnarray}
%L_{c,x} = \frac{\hbar}{p}\frac{1}{\sigma_{\theta_x}},\label{eqn:lcx}
%\end{eqnarray}
%where $p$ is the average electron beam momentum and $\sigma_{\theta_x}$ is the rms transverse angular spread of beam in one of the transverse direction perpendicular to the beam line ($z$) axis.   In the paraxial approximation, $\theta_x \approx p_x/p$ and the coherence length becomes:
%\begin{eqnarray}
%L_{c,x}  \approx \frac{\hbar}{\sigma_{p_{x}}}  =  \frac{\lambdabar_e}{\sigma_{\gamma\beta_{x}}}%,\label{eqn:lcx2}
%\end{eqnarray}
where $\lambdabar_e\equiv\hbar/m_ec =3.862...\times10^{-4}$ nm is the \emph{reduced} Compton wavelength of the electron. In this and all subsequent expressions, all fields and particle distributions are assumed symmetric about the beam line ($z$) axis.  %Both Eqns~(\ref{eqn:lcx}) and (\ref{eqn:lcx2}) make clear that a beam going through a waist will maximize the coherence length.  
Rewriting the coherence length in terms of the (normalized) emittance $\epsilon_{n,x}$ gives
\begin{eqnarray}
\frac{L_{c,x}}{\lambdabar_e}  = \frac{1}{\sigma_{\gamma\beta_{x}}} = \frac{\sigma_x}{\sqrt{\epsilon_{n,x}^2+\langle x\cdot\gamma\beta_{x}\rangle^2}}.%\\
%\left.\frac{L_{c,x}}{\lambdabar_e}\right|_{\mathrm{waist}} & = & \frac{\sigma_x}{\epsilon_{n,x}}
\end{eqnarray}
For a beam passing through a waist this expression reduces to \cite{ref:ued:dcbun2solconcept,ref:coldatoms1}
\begin{eqnarray} 
\left.\frac{L_{c,x}}{\lambdabar_e}\right|_{\mathrm{waist}} = \frac{\sigma_x}{\epsilon_{n,x}}.
\label{eqn:lcxemit}
\end{eqnarray} 
To determine how this quantity scales with the critical initial beam parameters and accelerating field requires relating the initial and final emittances.  Factoring out any emittance degrading effects occurring during transport allows one to write the emittance as: $\epsilon_{n,x,i} = f_{\epsilon} \cdot \epsilon_{n,x}$ where the factor $f_{\epsilon}\in(0,1]$ determines the degree of emittance preservation.  In general, $f_{\epsilon}$ depends strongly on the space charge dynamics along the beamline, which in turn are determined by the initial and final required beam sizes.  Nonetheless, using this and the expression for the emittance at the cathode $\epsilon_{n,x,i}=\sigma_{x,i}\sigma_{\gamma\beta_{x,i}}$, the coherence length can be written in terms of the magnification $M = \sigma_x/\sigma_{x,i}$ from cathode to sample as well as the initial coherence length:
\begin{eqnarray}
\frac{L_{c,x}}{\lambdabar_e}  = f_{\epsilon} \frac{\sigma_x}{\epsilon_{n,x,i}} = f_{\epsilon}M\cdot\frac{L_{c,x,i}}{\lambdabar_e}.\label{eqn:CLfactors1}
\end{eqnarray}
%It follows that optimal final coherence length requires simultaneously preserving emittance, as well as maximization of both the cathode to sample magnification and initial coherence length.  
The mean transverse energy (MTE) of the emitted electrons determines the initial coherence length \cite{ref:coldatoms1}:
\begin{eqnarray}
\frac{L_{c,x,i}}{\lambdabar_e}  =  \frac{1}{\sigma_{\gamma\beta_{x,i}}} = \sqrt{\frac{m_ec^2}{MTE}},
\end{eqnarray}
%and is thus independent of the beam dynamics.  However, for the charge densities under consideration here, space charge renders both $f_{\epsilon}$ and $M$ in Eqn~(\ref{eqn:CLfactors1}) strongly dependent on the initial 6D particle distribution rms sizes and shape, as well as the optics layout used to transport the beam to the sample.  
while the charge saturation limit, set by the desired extractable charge and cathode field, determines the size of the laser pulse.  Following \cite{ref:maxbb,ref:pietro2}, we write the aspect ratio of the photoemitted beam as $A = \sigma_{x,i} / \Delta z \approx \sigma_{x,i} /  \frac{eE_0}{m_ec^2}(c\sigma_{t,i})^2$, where $\Delta z \propto \frac{eE_0}{m_ec^2}(c\sigma_{t,i})^2$ gives the approximate length of the beam at the time of emission in terms of the field at the cathode $E_0$ and the laser pulse length $\sigma_{t,i}$.  In the charge saturation limit, this yields:
\begin{eqnarray}
\sigma_{x,i}\propto
\left\{ \begin{array}{c}
 (q/E_0)^{1/2}, \hspace{0.1cm} A \gg 1 \hspace{0.1cm}(``\mathrm{pancake}")\\
(q/\sigma_{t,i})^{2/3}E_0^{-1}, \hspace{0.1cm} A\leq 1 \hspace{0.1cm} (\mathrm{``cigar"})
\end{array} \right.
\end{eqnarray} 
Thus, the coherence length scales as:
\begin{eqnarray}
\frac{L_{c,x}}{\lambdabar_e} \propto  f_{\epsilon}\sigma_x \sqrt{\frac{m_ec^2}{MTE}}.
\left\{ \begin{array}{c}
 (E_0/q)^{1/2}, \hspace{0.1cm} A \gg 1 \\
E_0(\sigma_{t,i}/q)^{2/3}, \hspace{0.1cm} A\lesssim 1 
\end{array}. \right.
\label{eqn:lcxscale}
\end{eqnarray}
For beams with a large degree of emittance preservation, $f_{\epsilon}\approx 1$, and the above expression gives the correct scaling \cite{ref:maxbb,ref:pietro2}.

\section{One Approach for Optimal Coherence Length}

\begin{figure}[h!]
 \centering
\includegraphics[width=0.42\textwidth]{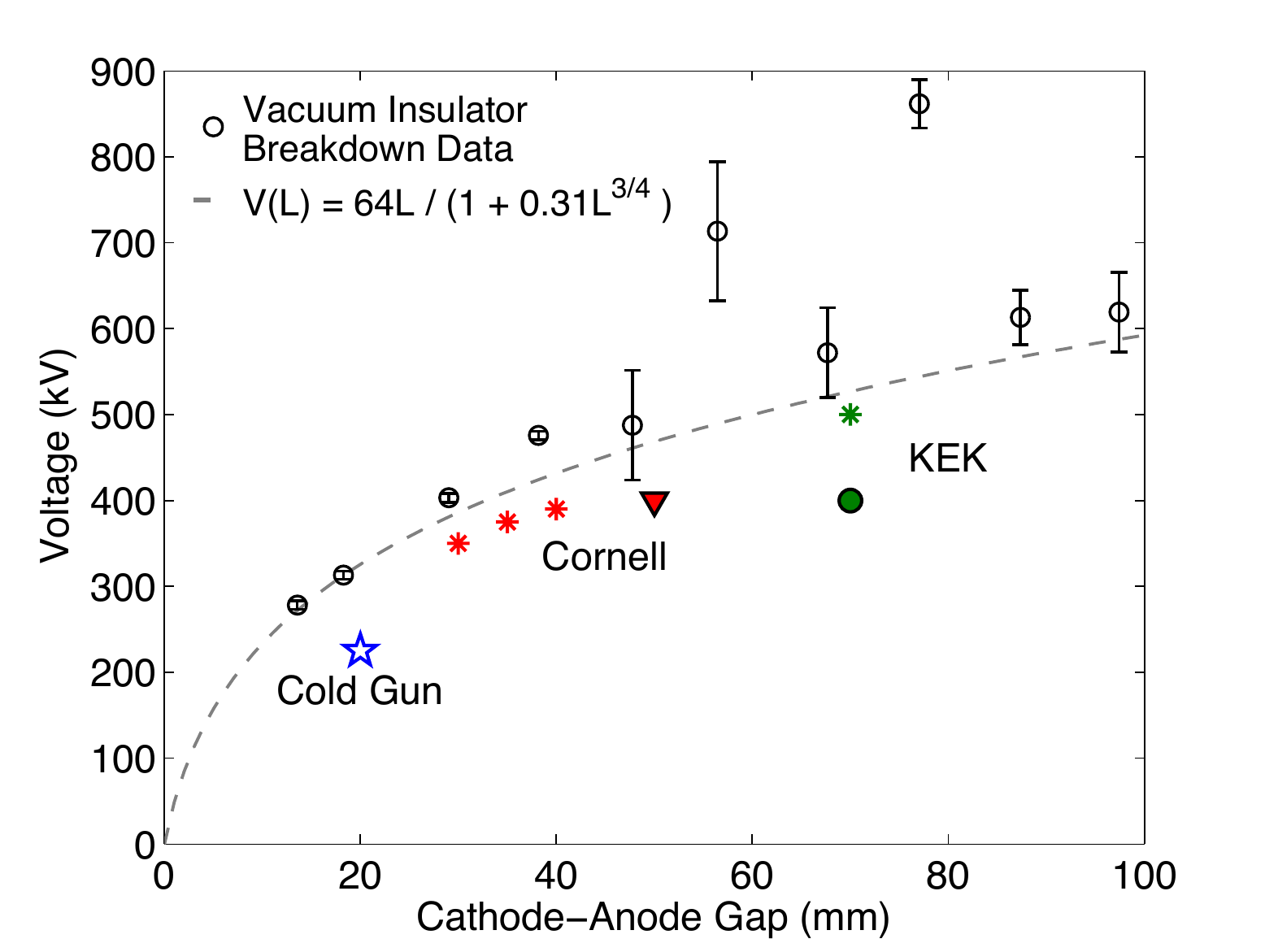}
 \caption{Voltage performance of high voltage DC systems as a function of the cathode-anode gap: (black) vacuum insulator breakdown data.  Additionally, the proposed gap and voltage for the Cornell Cryogun, (red) the stable processing voltages and gaps for the Cornell segmented gun with moveable anode with 300 pC bunches demonstrated at 400 kV (triangle), processing results of the second generation segmented gun at KEK (green), and voltage for beam tests (circle).}
\label{fig:voltage_vs_gap}
\end{figure}

Both limits in Eqn.~(\ref{eqn:lcxscale}) make clear that given a desired final spot size $\sigma_x$, and charge $q$ at the sample, maximizing coherence length requires larger cathode fields as well as smaller MTEs.  In this work, we seek to document the best coherence length achievable from photogun systems delivering the best in MTE technology.  %Ultimately, this leads to a comparison between gun technologies delivering the highest cathode fields, namely NCRF guns, and technology producing the best vacuums levels capable of supporting the lowest MTE cathodes, namely DC guns.  
To that end, we simulate a DC gun set-up, derived from the design of a 250 kV DC gun featuring a 20 mm cathode-anode gap, and  a  novel cryo-cooled photocathode system capable of cooling multi-alkali cathodes to 20 K under design and construction at Cornell University.  For this work, we specify the same gun geometry and a slightly lower gun voltage of 225 kV, in part guided by the empirical data on voltage breakdown and previous voltage demonstration figures for DC guns shown in Fig.~\ref{fig:voltage_vs_gap} \cite{ref:jaredRSI}.
Recently alkali antimonide photocathodes cooled to 90 K produced MTEs as low as a 15 meV \cite{ref:coldcathode}.  We anticipate MTEs of a few meV may be achievable in the new cryogun system, and thus, for simplicity, we assume a cathode MTE of 5 meV for all simulations for this beamline.

To model the gun fields, we use an analytic expression for the potential of a plate conductor with a hole in it immersed in a constant background field.    For this system, the potential is approximately:
\begin{eqnarray}
\Phi(r,z) &=& -E_{0}\left(\frac{R}{\pi}\right)\left[\sqrt{\frac{\rho-\lambda}{2}}-\frac{|z-L|}{R}\tan^{-1}\sqrt{\frac{2}{\rho+\lambda}}\right],
\nonumber
\\
\rho(r,z) &=& \sqrt{\lambda(r,z)^2 + 4(z-L)^2/R^2},
\nonumber
\\
\lambda(r,z) &=& \frac{1}{R^2}[(z-L)^2+r^2]-1.
\end{eqnarray}
In this expression, $E_0$ is the field at the cathode.  This solution becomes exact in the limit that the cathode-anode gap goes to infinity, and remains a good approximation provided that the radius of the anode hole is much greater than the gap ($R/L \ll 1$).  Here, the radius of the anode hole is 2.5 mm (compared to 20 mm for the gap), and the relative voltage error across the cathode is $<$ 1\%:
\begin{eqnarray}
\left|\frac{\Phi(z=0,r=0)}{\Phi(z=0,r=\infty)}-1\right| < 0.01.
\label{eqn:optgap} 
\end{eqnarray}
For this field set-up, the 225 kV gun voltage corresponds to a roughly 11 MV/m accelerating field at the cathode.  

\begin{figure}[h!]
 \centering
\includegraphics[width=0.42\textwidth]{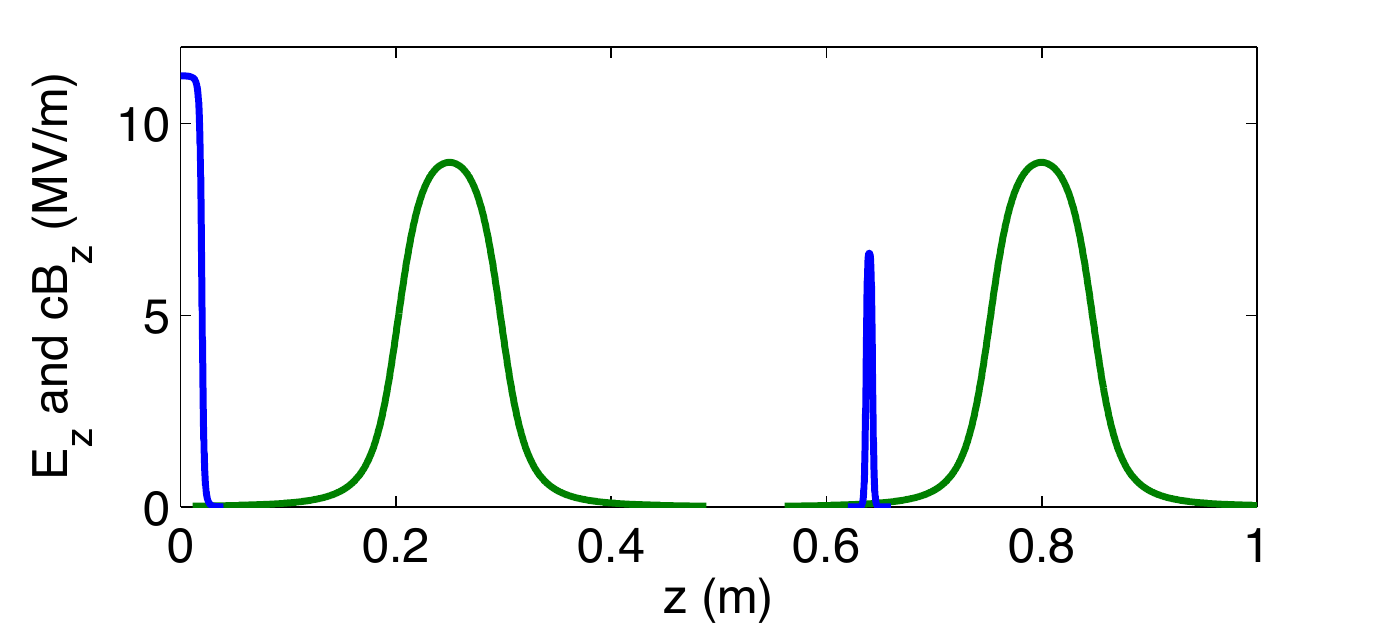}
 \caption{Example of the on-axis accelerating and solenoid field profiles for the cryogun set-up.}
\label{fig:cgunlayout}
\end{figure}

Fig.~\ref{fig:cgunlayout} shows the overall layout of the cryogun beamline.  This setup features a 3 GHz normal conducting buncher cavity for final bunch compression, as well as two solenoid magnets \cite{ref:ued:dcbun2solconcept,ref:ued:dcbun2sol2,ref:ued:dcbun2sol}.  For the buncher fields, we used the same 3 GHz field map as the Eindhoven design \cite{ref:ued:dcbun2solconcept} (a new 3 GHz design is currently underway at Cornell).  The solenoid field maps were created by scaling down the existing Cornell photoinjector fields by a factor of two.  We then fit the analytic form for the on-axis solenoid field from a sheet of current with radius $R$ and length $L$,
\begin{eqnarray}
B_z(z) = B_0\left( \frac{\Delta z_+}{\sqrt{\Delta z_+^2 + R^2}} -  \frac{\Delta z_-}{\sqrt{\Delta z_-^2 + R^2}}\right), 
\label{eqn:solfield}
\end{eqnarray}
where $\Delta z_{\pm} =  z \pm L/2$, to the solenoid field maps, and created a custom GPT element featuring the analytic result of the off-axis expansion of Eq.~(\ref{eqn:solfield}) to third order  in the radial offset $r$.  We note here that given the small beam sizes along each set-up ($\sigma_x \lesssim 2$ mm) determined by the optimizer, the first-order expansion of the solenoid fields accurately describes the beam dynamics through both beamlines.  Additionally, use of such small MTE values requires estimating the effect of disorder induced heating (DIH) near the cathode.  This issue is discussed later in the results section.

%Shaping of the laser is achieved by using the product of four archetypal distributions, shown in Fig.~\ref{fig:shapes} \cite{ref:dcrfcomp}.  For each shape type, there is an adjustable parameter ([0,1] for ``Tail", ``"Dip" and ``Ellipse" and [-1 1] for ``Slope") which controls the degree to the effect of the shape.  The same function is used to generate both the transverse and temporal laser shapes.  Examples are seen in Fig.~\ref{fig:xyshape} and Fig.~\ref{fig:tshape}.  In addition to varying the transverse and longitudinal laser shapes, the optimizer also selects the rms sizes.
%\begin{figure*}[ht!]
%    \begin{center}
%        \subfigure[\hspace{0.2cm}Archetypal laser shapes used to create arbitrary transverse and temporal laser shapes.]{%%
%            \label{fig:shapes}
%            \includegraphics[width=0.65\textwidth]{figs/shapes}
%        }\\ %  ------- End of the first row ----------------------%
%         \subfigure[\hspace{0.2cm}Example of a possible transverse laser distribution.]{%
%            \label{fig:xyshape}
%            \includegraphics[width=0.315\textwidth]{figs/xyshape}
%        }%
%        \subfigure[\hspace{0.2cm}Example of a possible temporal laser distribution.]{%
%           \label{fig:tshape}
%           \includegraphics[width=0.33\textwidth]{figs/tshape}
%        }\\ %  ------- End of the first row ----------------------%
%    \end{center}
%    \caption{%
%    \label{fig:3Dslice}
%    Generation of arbitrary transverse and temporal laser shapes: (a) 4 archetypal distributions, (b-c) example distributions.
%     }%
%\end{figure*}

\section{Results}

In order to produce the best coherence length performance from the cryogun UED setup, multi-objective genetic optimizations were performed using General Particle Tracer and the same optimization software used previously in \cite{ref:lowemitter,ref:lowemitter2,ref:lowemitter3}.  For these simulations, the optimizer varied the laser rms sizes, beamline element parameters and positions.  Additionally, the optimizer was allowed to arbitrarily shape both the transverse and longitudinal laser distributions, based on the same method described in \cite{ref:dcrfcomp}.  Table-\ref{tab:beamline_params} displays the beamline parameters varied for each setup.
\begin{table}[htb]
\caption{Beamline Simulation Parameters}\label{tab:beamline_params}
\begin{ruledtabular}
\begin{tabular}{ cc | cc }
Parameter & & Range  \\ 
\hline
Initial Charge & &[0,1000]  fC\\
Laser Size $\sigma_{t,i}$ &&  [0,20] ps \\
Laser Size $\sigma_{x,i}$ && [0,1] mm  \\
Cathode MTE && 5 meV  \\
Peak Gun Field && 11.1 MV/m \\
Solenoid 1 Peak Field && [0, 0.48] T\\
Solenoid 1 Position && [0.17,  0.67]  m\\
Peak Buncher Field && [0.0, 15] MV/m \\
Buncher Phase && [0, 360] deg \\
Buncher Position && [0.27, 1.27] m \\
Solenoid 2 Peak Field && [0.0, 0.48] T \\
Solenoid 2 Position && [0.37, 1.87]  m\\
Sample Position && [0.37, 3.87] m
\end{tabular}
\end{ruledtabular}
\end{table}

\subsubsection{Optimal Emittance}

\begin{figure*}[htb!]
    \begin{center}
        %\ %  ------- End of the first row ----------------------%
         \subfigure[\hspace{0.2cm}Optimal emittance as a function of bunch charge.]{%
            \label{fig:evsq}
            \includegraphics[width=0.42\textwidth]{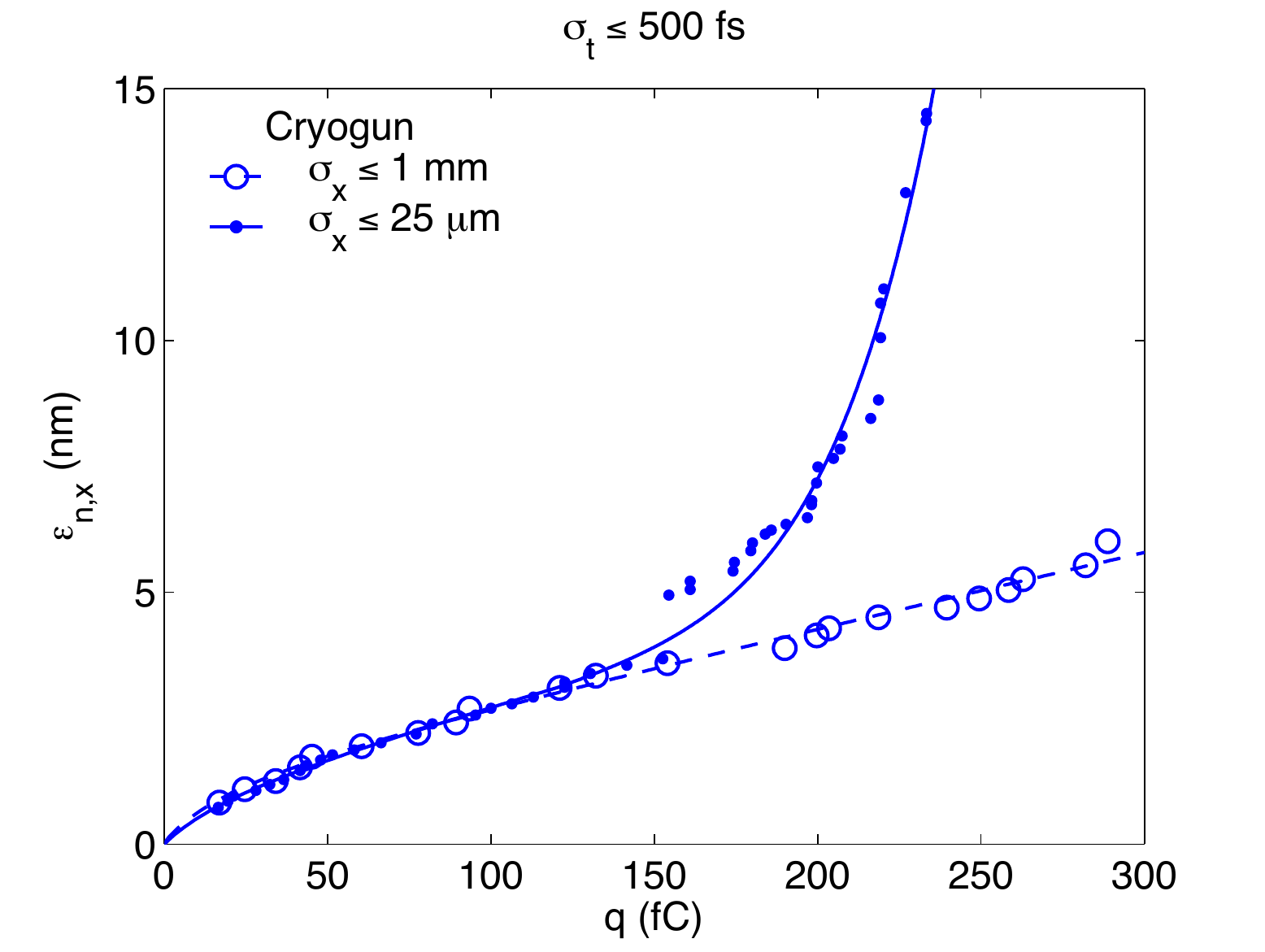}
        }%
        \subfigure[\hspace{0.2cm}Optimal emittance as a function of bunch charge.]{%
           \label{fig:evsqscaling}
           \includegraphics[width=0.42\textwidth]{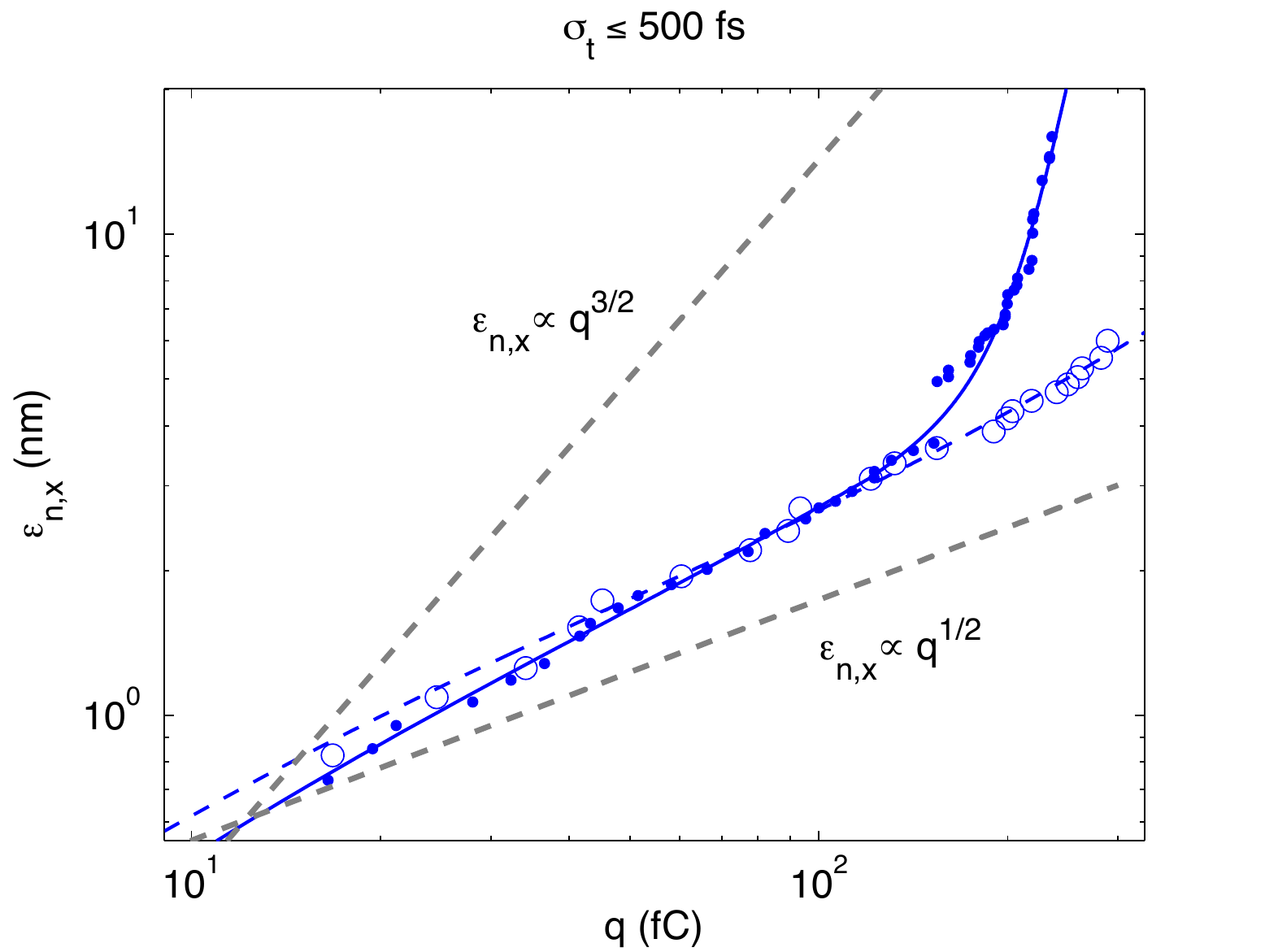}
        }\\ %  ------- End of the first row ----------------------%
    \end{center}
    \caption{%
    \label{fig:EvsQ}
    The optimal emittances for different final tranverse spot sizes. Dashed lines indicate a final beam size of $\sigma_x\leq1$ mm, while solid lines indicate a final beam size of $\sigma_x\leq25$ $\mu$m. (b) Shows the same data on a log-log plot.   Dashed grey lines indicate how the emittance should scale with $q$ based on Eqn.~(\ref{eqn:lcxscale}).
     }%
\end{figure*}

Given a final spot size $\sigma_x$ Eqn.~(\ref{eqn:lcxemit}-\ref{eqn:lcxscale}) imply the fundamental limit to the coherence is the emittance at the sample.   As previously stated, the emittance preservation factor $f_{\epsilon}$ in Eqn.~(\ref{eqn:lcxscale}) determines the degree to which the scaling laws in this expression hold true, and may depend strongly on both the initial and final beam sizes.  To determine the effects of constraining the final required rms sizes, we perform an initial round of optimizations for a ``large" final beam, $\sigma_x\leq1$ mm and $\sigma_t\leq 500$ fs, and compare that to optimizations with the smallest final spot size considered in this work, $\sigma_x\leq25$ $\mu$m.  In these optimizations, we require that no particles are lost in beam transport (later we allow for clipping of the beam at the sample).  Fig.~\ref{fig:evsq} shows the emittance performance for both spot sizes.  In these and all similar plots, we fit a rational polynomial to the Pareto front in order to better guide the eye and to aid estimating and interpolating between points on the front.  As the data shows, the emittance performance for both final beam sizes is similar at lower charges.  In the case of the 25 $\mu$m spot size, the emittance suffers for charges above roughly 150 fC, as the space charge repulsion makes focusing/compressing the bunch down to the desired final beam sizes more difficult.
Fig.~\ref{fig:evsqscaling} shows the same data on a log-log plot.  The grey dashed lines represent the scaling of the emittance with charge predicted by Eqn.~(\ref{eqn:lcxscale}).  Computing the the initial beam aspect ratio $A$ for each front yields 0.1 - 0.2.   Note that the emittance scales as $q^{1/2}$ up to roughly $150$ fC, though the aspect ratio indicates its operation in the long beam regime.  

%In order to get a sense to the scale of for the coherence length for different bunch charges, we estimate the coherence length for a final beam size of 25 $\mu$m using the fitted emittance curves in Fig.~\ref{fig;evsq}.  Fig.~\ref{fig:lcxestimate} shows the results.  Note that the dashed lines estimate the coherence for $\sigma_x\leq25$ $\mu$m if one was able to produce the emittance values obtained with $\sigma_x\leq25$, while the solid lines use the emittance actually obtained at $\sigma_x\leq 25$ $\mu$m.  
%\begin{figure}[htb]
% \centering
%\includegraphics[width=90mm]{figs/comp_Lcx_q}
% \caption{Estimate of the coherence length with $\sigma_x\leq$ 25 $\mu$m.}
%\label{fig:lcxestimate}
%\end{figure}
%Using the data in the solid lines leads to coherence lengths of 2.03 (3.64) nm for $10^5$ electrons and 11.4 (12.7) nm for $10^6$ electrons for the cryo-gun (rf gun), respectively.  This curve defines the worst emittance performance for each gun type tested thus far, and therefore sets lower bound for the coherence length at other final spot sizes in the limit of a long final pulse length ($\sigma_t\leq500$ fs).

\subsubsection{Optimal Coherence Length}

From the emittance vs. charge data in Fig.~\ref{fig:EvsQ}, we selected solutions corresponding to $10^5$ and $10^6$ electrons at the sample and seeded a new set of optimizations maximizing the coherence length and minimizing the final bunch length at the sample $\sigma_t$.  The inclusion of a pinhole representing the sample allowed the optimizer to now clip particles at the sample location, subject to the constraint that $q_f \geq 10^5$ or $q_f \geq 10^6$ electrons after particle clipping at the iris.  For each bunch charge, optimizations were first run with the smallest sample radius  $R=50$ $\mu$m.  These optimizations provided the initial seed for simulations with with $R=100$ $\mu$m, as the results for the smallest pinhole automatically satisfy all of the contraints for the next larger pinhole.  Likewise, the optimization results with $R=100$ $\mu$m provided viable solutions to seed simulations with $R=200$ $\mu$m.  
\begin{figure*}[ht!]
    \begin{center}
        %\ %  ------- End of the first row ----------------------%
         \subfigure[\hspace{0.2cm}Optimal coherence length vs bunch at the sample.]{%
            \label{fig:LvsT_1e5}
            \includegraphics[width=0.42\textwidth]{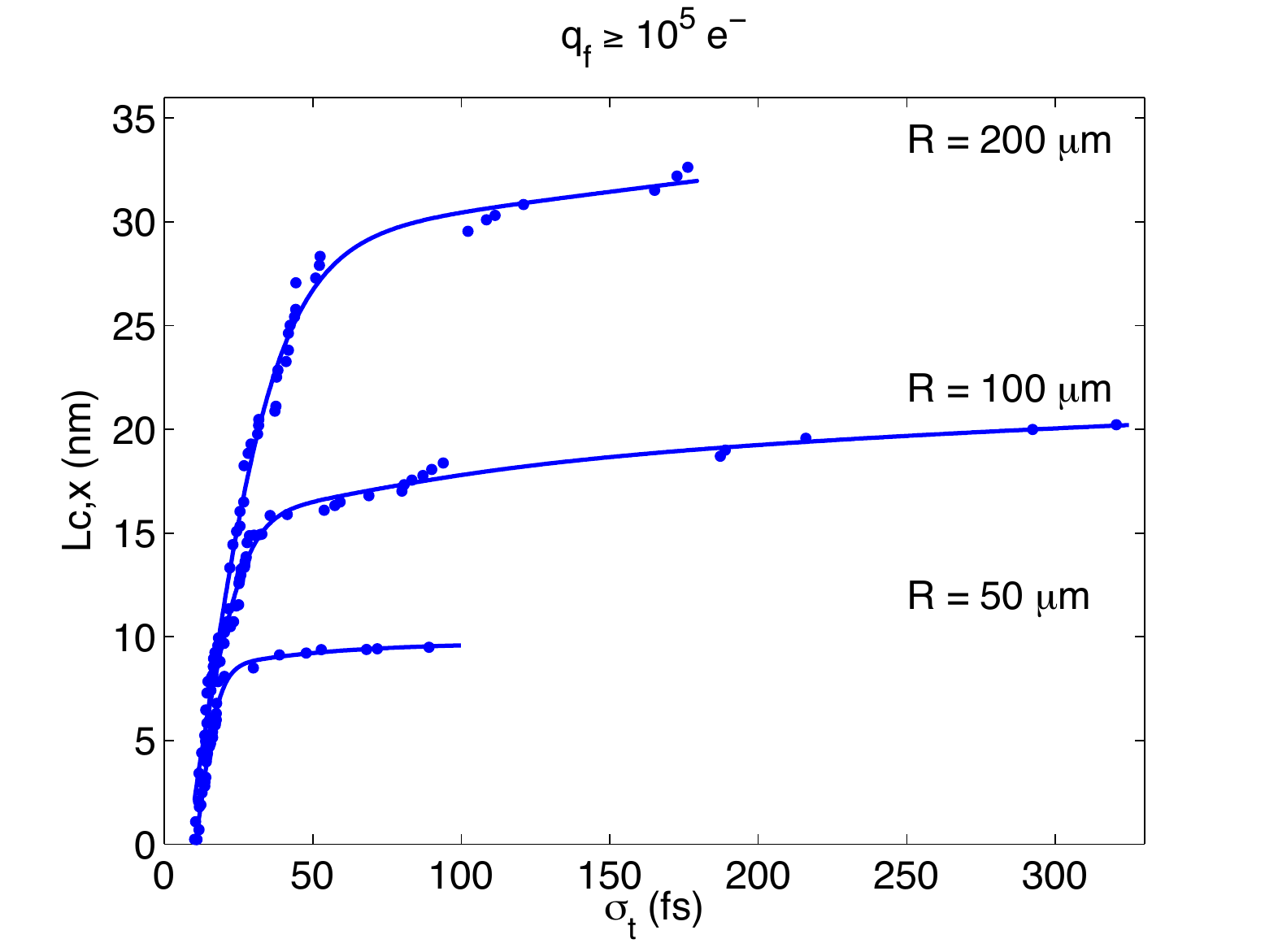}
        }%
        \subfigure[\hspace{0.2cm}Optimal coherence length vs bunch at the sample.]{%
           \label{fig:LvsT_1e6}
           \includegraphics[width=0.42\textwidth]{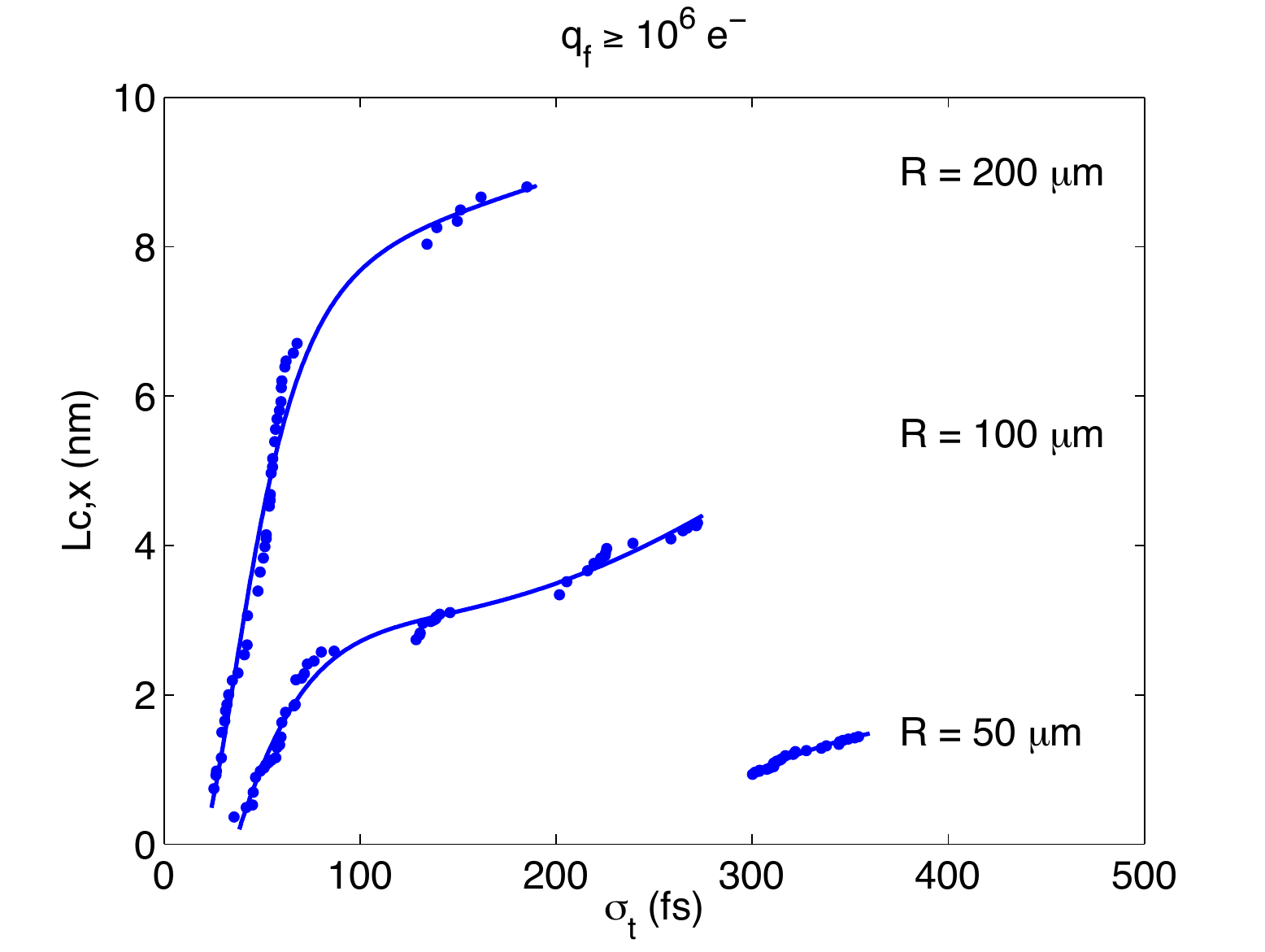}
        }\\ %  ------- End of the first row ----------------------%
    \end{center}
    \caption{%
    \label{fig:eLvsT}
    Optimal coherence length as a function of bunch length with charge on the sample of (a) $10^5$ and (b) $10^6$ electrons.  
     }%
\end{figure*}
 For all simulations, 6k macro-particles were used, and the initial charge was allowed to vary up to 1 pC, which implies that at least 100 macroparticles must survive the clipping at the sample for the smallest final allowed charge of $q_f \geq 10^5$ electrons.  Upon producing the optimum fronts, additional simulations were run with 30k macro-particles to check the statistics after clipping, and reproduced the coherence lengths computed with 6k initial macroparticles to within 20\%.

Fig.~\ref{fig:eLvsT} shows the optimal coherence length as a function of final bunch length $\sigma_t$ for each bunch charge and sample radius.  For $q_f \geq 10^5$ electrons, the cryogun beamline provides solutions with $\sigma_t\lesssim 100$ fs for all three pinhole sizes.  Computing the relative coherence length ($L_{c,x}/\sigma_x)$  for a final bunch length of $\sigma_t\approx 100$ fs using the data from the fits to the optimization results (solid lines) and approximating $\sigma_x\approx R/2$ gives 0.27 nm/$\mu$m.  %We note the ratio of these two results is 1.15, remarkably similar to the estimate in Eqn.~\ref{eqn:Lratio}.
Increasing the required final charge to $q_f \geq 10^6$ electrons produces more varied coherence performance.  For final spot sizes of $\sigma_x\geq 50$ $\mu$m and final bunch lengths of $\sigma_t\approx200$ fs, the cryogun beamline produces a relative coherence length of roughly 0.11 nm/$\mu$m.  For these parameters, estimating the relative coherence length gives 0.1 nm/$\mu$m for a final $\sigma_t\leq100$ fs.  Table-\ref{tab:LcxOpt} summarizes these values.
%\begin{table}[h!]
%\caption{Relative Coherence Length Values (nm/$\mu$m)}\label{tab:LcxOpt}
%\begin{ruledtabular}
%\begin{tabular}{ c | c c c }
%Beamline & Cryogun& NCRF \\ 
%\hline
%$q_f\geq10^5$ $\mathrm{e}^-$, $\sigma_x\geq25$ $\mu$m, $\sigma_t\approx100$ fs & 0.27 & 0.31  \\
%$q_f\geq10^6$ $\mathrm{e}^-$, $\sigma_x\geq50$ $\mu$m, $\sigma_t\approx100$ fs  & 0.10 & NA \\
%$q_f\geq10^6$ $\mathrm{e}^-$, $\sigma_x\geq50$ $\mu$m, $\sigma_t\approx200$ fs  & 0.11 & 0.06  
%\end{tabular}
%\end{ruledtabular}
%\end{table}
\begin{table}[h!]
\caption{Relative Coherence Length Values (nm/$\mu$m)}\label{tab:LcxOpt}
\begin{ruledtabular}
\begin{tabular}{ c | c c }
Beamline & $L_{c,x}/\sigma_x$ \\ 
\hline
$q_f\geq10^5$ $\mathrm{e}^-$, $\sigma_x\geq25$ $\mu$m, $\sigma_t\approx100$ fs & 0.27   \\
$q_f\geq10^6$ $\mathrm{e}^-$, $\sigma_x\geq50$ $\mu$m, $\sigma_t\approx100$ fs  & 0.10  \\
$q_f\geq10^6$ $\mathrm{e}^-$, $\sigma_x\geq50$ $\mu$m, $\sigma_t\approx200$ fs  & 0.11   
\end{tabular}
\end{ruledtabular}
\end{table}
If the coherence length (considering only the dynamics of the inner portion of the beam that survives clipping) scales as $q_f^{-\nu}$, then the ratio of the two required final charges for the curves in Fig.~\ref{fig:LvsT_1e6} and \ref{fig:LvsT_1e6} implies $\nu=\log_{10}(L{c,x}(10^5\hspace{0.1cm}\mathrm{e}^-)/L{c,x}(10^6\hspace{0.1cm}\mathrm{e}^-))$.  Roughly estimating the coherence length ratios from the asymptotic portions of the solid curves in Fig.~\ref{fig:LvsT_1e6} and \ref{fig:LvsT_1e6} gives $\nu=$ 0.76-0.81, 0.53-0.6, and 0.55-0.59 for the $R=50$, 100, and 200 $\mu$m curves respectively.  This suggests that the coherence length data may roughly scale as $q^{-1/2}$ for the larger two of the three sample radii.

In addition to determining the the optimal coherence length, the optimizations producing the data in Fig.~\ref{fig:eLvsT} also provides information about the optimal positioning of the beamline elements in each set-up.
\begin{figure*}[ht!]
    \begin{center}
        %\ %  ------- End of the first row ----------------------%
         \subfigure[\hspace{0.2cm}Cryogun setup beamline element positions.]{%
            \label{fig:LvsT_1e5}
            \includegraphics[width=0.42\textwidth]{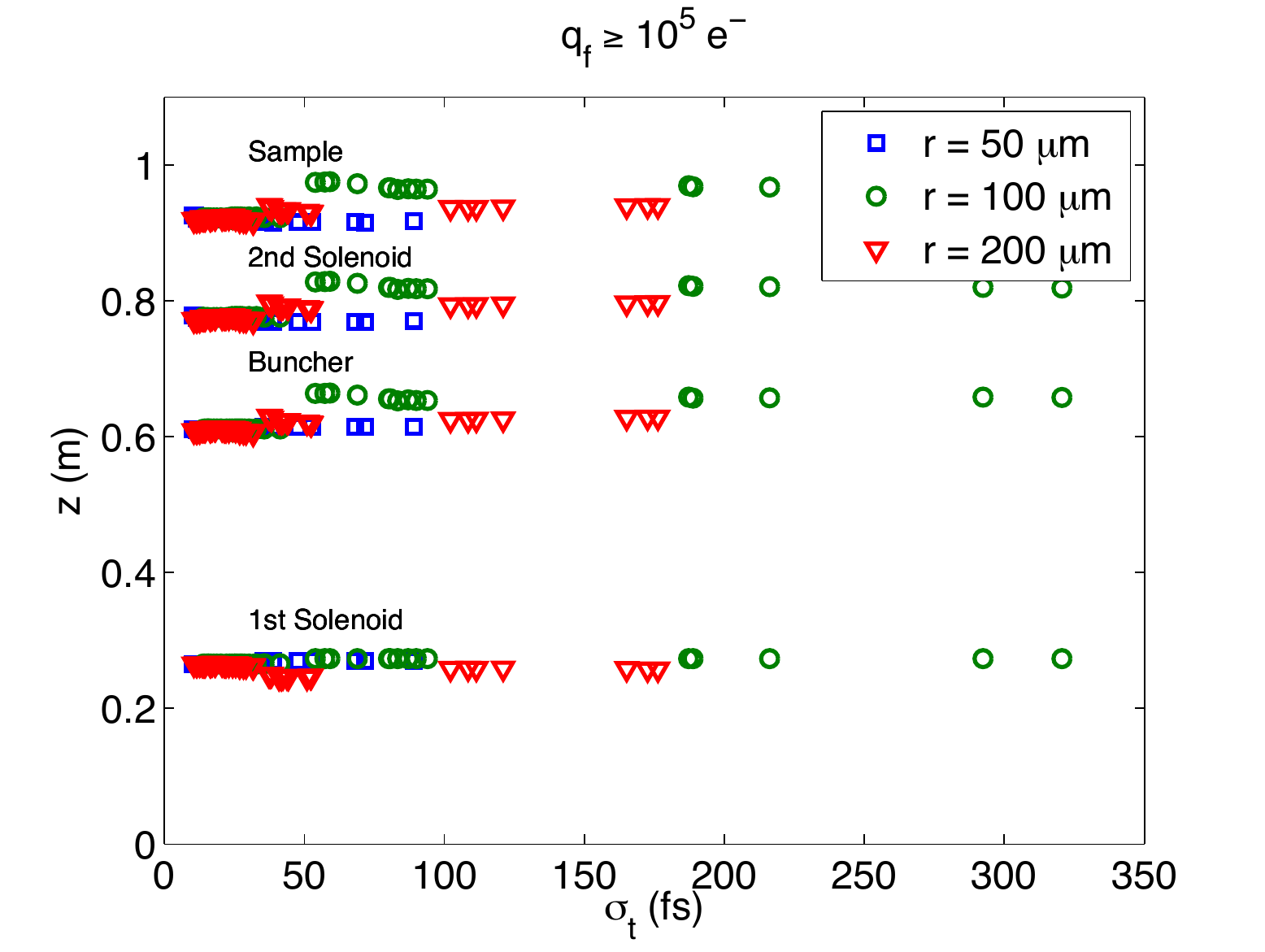}
        }%
        \subfigure[\hspace{0.2cm}Cryogun setup beamline element positions.]{%
           \label{fig:LvsT_1e6}
           \includegraphics[width=0.42\textwidth]{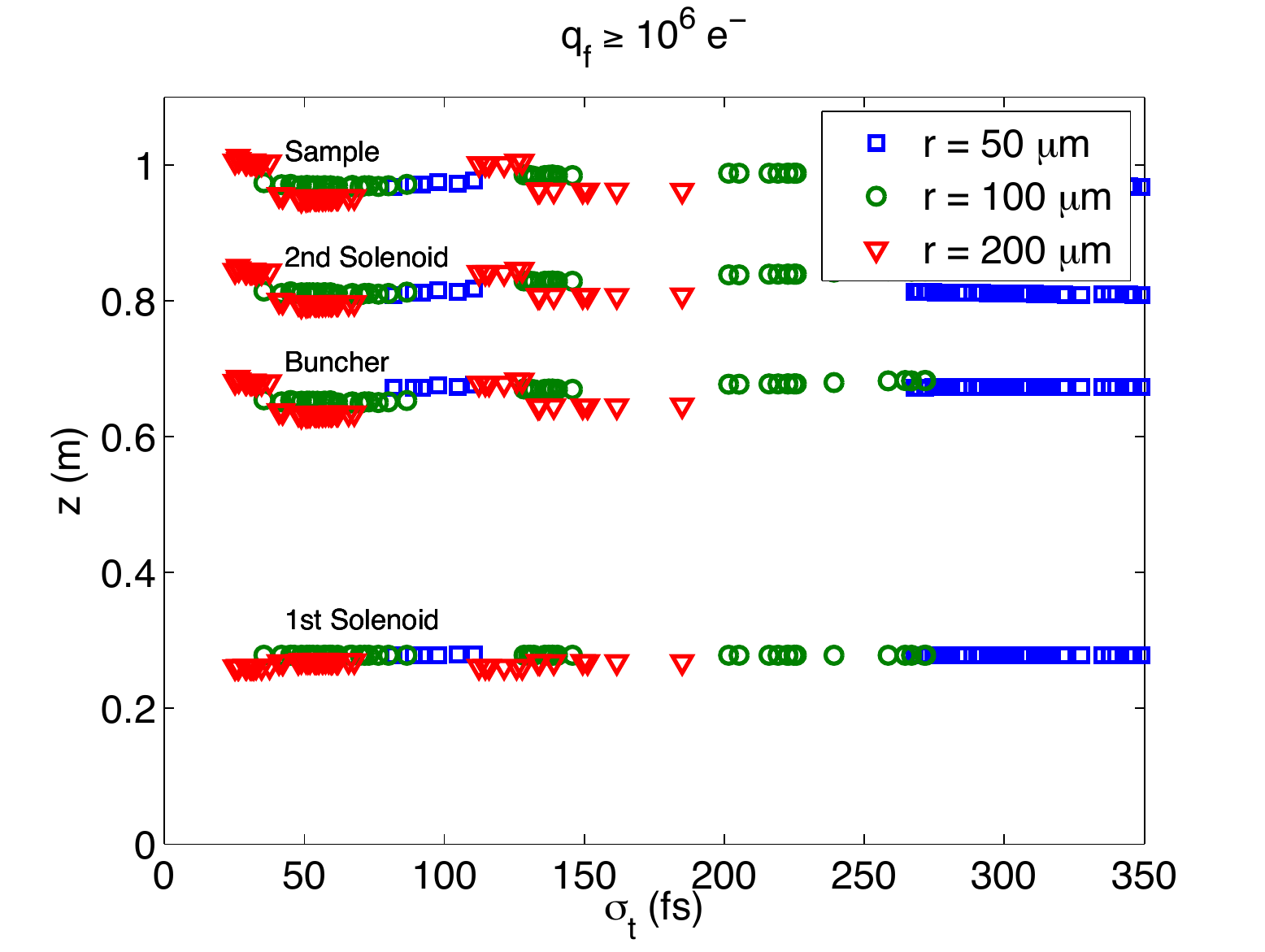}
        }\\ %  ------- End of the first row ----------------------%
        %\ %  ------- End of the first row ----------------------%
        % \subfigure[\hspace{0.2cm}NCRF setup beamline element positions.]{%
        %    \label{fig:LvsT_1e5}
        %    \includegraphics[width=0.42\textwidth]{figs/ncrf_zs_1e5}
        %}%
        %\subfigure[\hspace{0.2cm}NCRF setup beamline element positions.]{%
        %   \label{fig:LvsT_1e6}
        %   \includegraphics[width=0.42\textwidth]{figs/ncrf_zs_1e6}
        %}\\ %  ------- End of the first row ----------------------%
    \end{center}
    \caption{%
    \label{fig:elepos}
    Position of the beamline elements for a final charge at the sample of $10^ 5$ electrons (a) and $10^ 6$ electrons (b).     }%
\end{figure*}
Fig.~\ref{fig:elepos} shows the positions of the beamline elements corresponding to the optimizations shown in Fig.~\ref{fig:eLvsT}.  For the cyrogun beamline the optimizer eventually settled on fairly fixed element positions both final charges and all sample radii.  %The element positions chosen for the S-band gun beamline by the optimizer show a larger variation.  
Table-\ref{tab:beamline_params} displays the element positions averages over all the results of all six optimization shown in Fig.~\ref{fig:eLvsT}.
\begin{table}[htb]
\caption{Average Optimized Beamline Element Positions}\label{tab:beamline_params}
\begin{ruledtabular}
\begin{tabular}{ cc | cc  }
Element &&  Position  \\ 
\hline
Solenoid 1  && 0.27 m   \\
Buncher Cavity && 0.64 m \\
Solenoid 2 &&  0.80 m \\
Sample Pinhole  && 0.95 m   
\end{tabular}
\end{ruledtabular}
\end{table}
%We point out the location of the first solenoid is the same for both systems, and in the case of the RF gun, pushed to the minimum distance allowed from the cathode.

\subsubsection{Example Simulations}
In order to get a better feel of the beam dynamics determined by the coherence length optimizations, we ran several example solutions from the coherence vs. final bunch length fronts shown in Fig.~\ref{fig:eLvsT}.  From these, we present two examples, one for each of the final charges.  In all cases shown, the final sample radius was $R=100$ $\mu$m.  The final rms bunch lengths was set to $\sigma_t\approx$ 100 fs and 200 fs for the lower/higher final charge, respectively.  Table-\ref{tab:resparams} displays the resulting relevant beam parameters.
%\begin{table}[h!]
%\caption{Example results from optimal fronts for a sample radius of $R=100$ $\mu$m.}\label{tab:resparams}
%\begin{ruledtabular}
%\begin{tabular}{ c | c c  }
%$q_f\geq10^5$ $e^-$, $\sigma_{t,f}\sim100$ fs  & Cryogun & NCRF Gun  \\ 
%\hline
%Estimated DIH (meV) & 0.75 & 1.8 \\ 
%laser $\sigma_{x,y}$ ($\mu$m) & 5.36 & 8.06  \\
%laser $\sigma_t$ (fs) & 8280 & 550  \\
%Aspect Ratio A & 0.04 & 2.0 \\
%$q_i$ (fC) & 47.2 & 860  \\
%$q_f/q_i$ & 0.35 & 0.02  \\
%$\epsilon_{n,x}$ (nm) & 1.05 & 1.07\\
%$L_{c,x}$ (nm) & 18.1 & 17.8\\
%$\lambdabar_e\sigma_x/\epsilon_{n,x}$ (nm) & 18.4 & 18\\
%\hline
%\hline
%$q_f\geq10^6$ $e^-$, $\sigma_{t,f}\sim200$ fs & Cryogun & NCRF Gun \\
%\hline
%Estimated DIH (meV) & 1.6 & 12 \\ 
%laser $\sigma_{x,y}$ ($\mu$m) & 5.83 & 11  \\
%laser $\sigma_t$ (fs) & 7310 & 73.8  \\
%Aspect Ratio A & 0.06 & 145\\
%$q_i$ (fC) & 239 & 268  \\
%$q_f/q_i$ & 0.73 & 0.65  \\
%$\epsilon_{n,x}$ & 5.27 & 4.4  \\
%$L_{c,x}$ (nm) & 3.25 & 4.3 \\
%$\lambdabar_e\sigma_x/\epsilon_{n,x}$ (nm) & 3.7 & 4.4 \\
%\end{tabular}
%\end{ruledtabular}
%\end{table}

\begin{table}[ht!]
\caption{\label{tab:proj_emit}Data for examples with $10^5$ electrons, $R=100$, $\mu$m $\sigma_t\approx 200$ fs.}
\subtable[\hspace{0.2cm}Horizontal projected emittance data.]{
\label{tab:resparams}
\begin{ruledtabular}
\begin{tabular}{c | cccc}
Parameter & Cryogun  \\
\hline
Estimated DIH & 0.75 meV \\ 
laser $\sigma_{x,y}$ & 5.36 $\mu$m \\
laser $\sigma_t$ & 8280 fs \\
Aspect Ratio A & 0.04 \\
$q_i$ & 47.2 fC  \\
$q_f/q_i$ & 0.35 \\
$\epsilon_{n,x}$ & 1.05 nm \\
$L_{c,x}$ & 18.1 nm \\
$\lambdabar_e\sigma_x/\epsilon_{n,x}$ & 3.7 nm  \\
\end{tabular}
\end{ruledtabular}
}
\subtable[\hspace{0.2cm}Data for examples with $10^6$ electrons, $R=100$, $\mu$m $\sigma_t\approx 200$ fs.]{
\label{tab:proj_emit:y}
\begin{ruledtabular}
\begin{tabular}{c | cccc} 
Parameter & Cryogun  \\
\hline
Estimated DIH  & 1.6 meV  \\ 
laser $\sigma_{x,y}$  & 5.83  $\mu$m \\
laser $\sigma_t$ & 7310  fs\\
Aspect Ratio A & 0.06 \\
$q_i$ & 239  fC \\
$q_f/q_i$ & 0.73   \\
$\epsilon_{n,x}$ & 5.27  nm\\
$L_{c,x}$ & 3.25  nm \\
$\lambdabar_e\sigma_x/\epsilon_{n,x}$ & 3.7 nm \\
\end{tabular}
\end{ruledtabular}
}\\
\end{table}

\begin{figure*}[ht!]
    \begin{center}
        \subfigure[\hspace{0.2cm}]{%
           \label{fig:cgun_stdx}
           \includegraphics[width=0.95\textwidth]{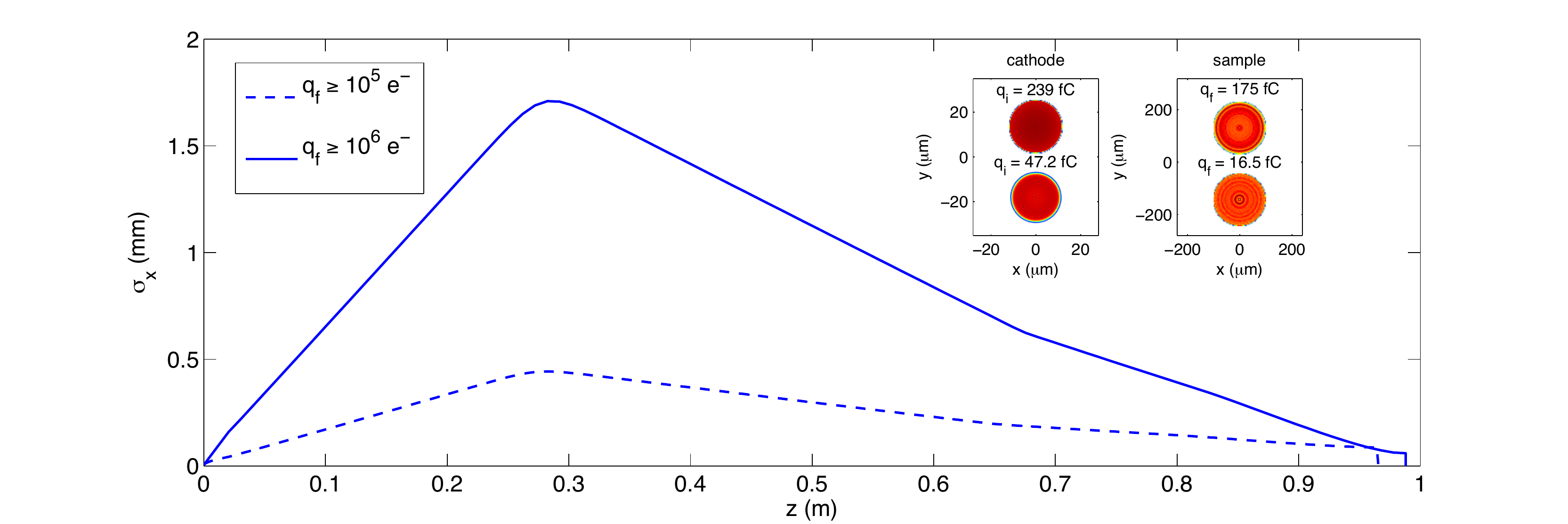}
        }\\ %  ------- End of the first row ----------------------%
        \subfigure[\hspace{0.2cm}]{%
           \label{fig:cgun_stdt}
           \includegraphics[width=0.95\textwidth]{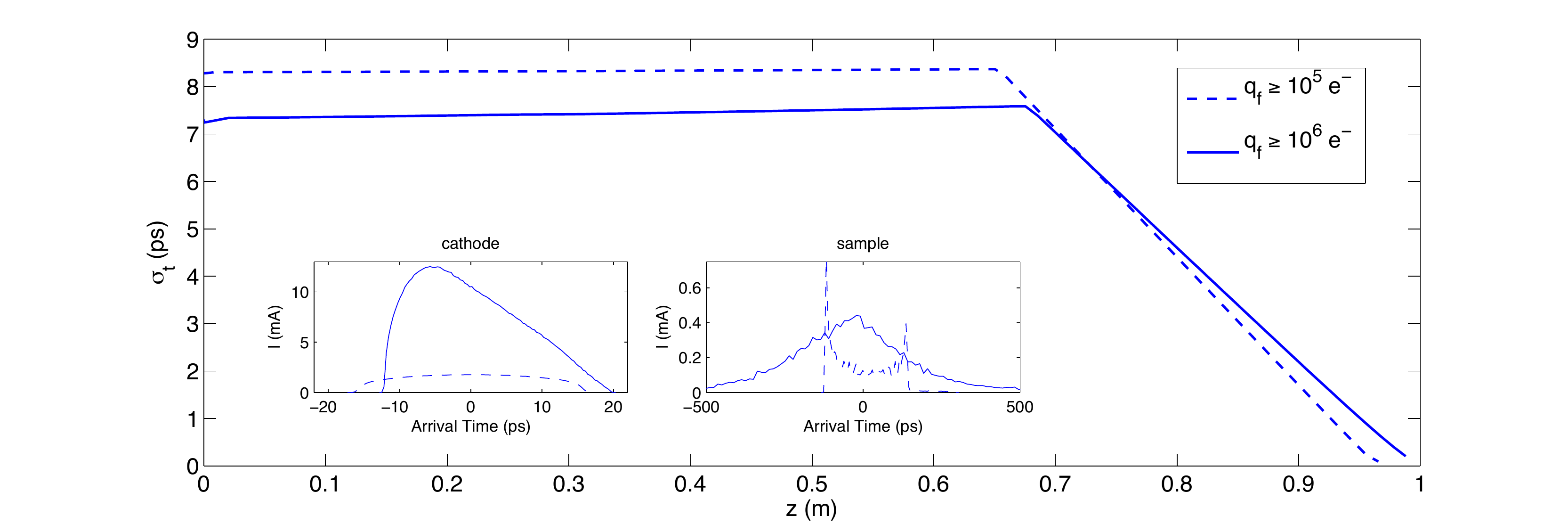}
        }\\ %  ------- End of the first row ----------------------%
    \end{center}
    \caption{%
    \label{fig:stdx}
    Transverse (a) and longitudinal (b) rms beam size along the cyrogun final charges of $10^ 5$ (dashed line) and $10^ 6$ electrons (solid line).  Insets show the transverse (a) and longitudinal (b) beam distributions at the cathode and sample locations for both final charges.}%
\end{figure*}

Fig.\ref{fig:cgun_stdx} shows the transverse rms beamsize along the cryogun beamline, as well as the initial transverse laser profile and the final electron beam transverse distribution at the sample for both bunch charges.  The optimizer chose a roughly flattop transverse laser profile with $\sigma_x\approx 5$ $\mu$m for both final charges.  The clipping at the sample produces a roughly flattop transverse electron beam distribution, validating the approximation $\sigma_x\approx R/2$ used to compute the relative coherence lengths in Table-\ref{tab:LcxOpt}.
The optimizer chose a smaller transmission $T=q_f/q_i$ for the smaller final charge $q_f\geq10^5$ electrons, with $T=35$\% transmission.  At $q_f\geq10^6$ electrons, the optimizer clipped fewer particles, resulting in a transmission of $T=73$\%.

Fig.~\ref{fig:cgun_stdt} shows the rms bunch length, and the initial temporal current profile produced by the laser, and the electron beam current profile at the sample.  The use of the buncher cavity allows for a fairly constant bunch length along the beamline up to the cavity, where the buncher applies an energy chirp which results in the bunch being compressed by the time it reaches the screen.
%\begin{figure*}[ht!]
%    \begin{center}
%        \subfigure[\hspace{0.2cm}]{%
%           \label{fig:cgun_stdt}
%           \includegraphics[width=0.95\textwidth]{figs/cgun_stdt}
%        }\\ %  ------- End of the first row ----------------------%
%        \subfigure[\hspace{0.2cm}]{%
%           \label{fig:ncrf_stdt}
%           \includegraphics[width=0.95\textwidth]{figs/ncrf_stdt}
%        }\\ %  ------- End of the first row ----------------------%
%    \end{center}
%    \caption{%
%    \label{fig:stdt}
%    Longitudinal rms beam size along the cyrogun (a) and NCRF beamlines (b) for final charges of $10^ 5$ (dashed line) and $10^ 6$ electrons (solid line).  Insets show the longitudinal current profile at the cathode (laser distribution) and sample locations for both final charges.}%
%\end{figure*}
%The lack of a separate buncher cavity for the RF gun beamline requires different dynamics.  Specifically, the laser pulse is significantly shorter at both charges.  The RF gun achieves further bunching by being run -43 (40) deg from the peak field at the cathode occurring at $t=0$.  This provides significant bunching for the lower charge example (dashed line). 
The optimizer chose a roughly flattop longitudinal initial laser distribution for the lower charge, and a sloped distribution at the higher charge.  %We note that at the larger charge, the RF gun produces a more uniform pulse than cryogun, which shows longer tails in the final current profile at the sample.

From the transverse and longitudinal rms data, the initial electron beam volume and aspect ratio follow, which allows for the estimation of the of disordered induced heating near the cathode surface, as well which scaling law regime from Eqn.~(\ref{eqn:lcxscale}) should apply to the dynamics.  In both cases, we assume a uniform beam with equivalent rms sizes.  From this, the volume follows:
\begin{eqnarray}
V &=& \pi R^2 L\approx \pi (2\sigma_x)^2 \cdot \frac{1}{2}\frac{eE_0}{m} (\sqrt{12}\sigma_t)^2 
\nonumber\\
&\approx& \frac{24\pi E_0}{mc^2 [\mathrm{eV}]}\sigma_x^2(c\sigma_t)^2
\end{eqnarray}
Using this to compute the electron number density for each of the example cases yields $4\times10^{17}$ ($4\times10^{18}$) $\mathrm{m}^{-3}$ for the final charges at the sample of $10^5$ ($10^6$) electrons respectively.  From this we estimate the effect of disordered induced heating using the formula given by Maxson: $\Delta kT \hspace{0.1cm}[\mathrm{eV}]  = 1.04\times10^{-9}(n_0 \hspace{0.1cm}[m^{-3}])^{1/3}$ \cite{ref:DIH}, where $n_0$ is the electron number density.  For the examples, this yields a DIH effect of 0.75 and 1.6 meV for the lower/higher final sample charge, or roughly 15\% and 32\% of the original 5 meV cathode MTE.  %Similarly, for the RF gun parameters, the DIH estimate gives 1.8 and 12 meV for the lower/higher final sample charge.  %This corresponds to roughly 5\% and 34\% of the initial 35 meV MTE for this set-up.  
Computing the initial electron beam aspect ratio yields $A = 0.04$ ($0.06$) with a final charge of $10^5$ ($10^6$) electrons, respectively.  As anticipated from the emittance optimizations, the cryogun produces best performance when operating in the long initial electron beam limit.%, while the RF gun, due to the lack of a separate buncher, operates best in the short pulse regime.

\begin{figure*}[ht!]
    \begin{center}
        \subfigure[\hspace{0.2cm}]{%
           \label{fig:enx_cgun1}
           \includegraphics[width=0.95\textwidth]{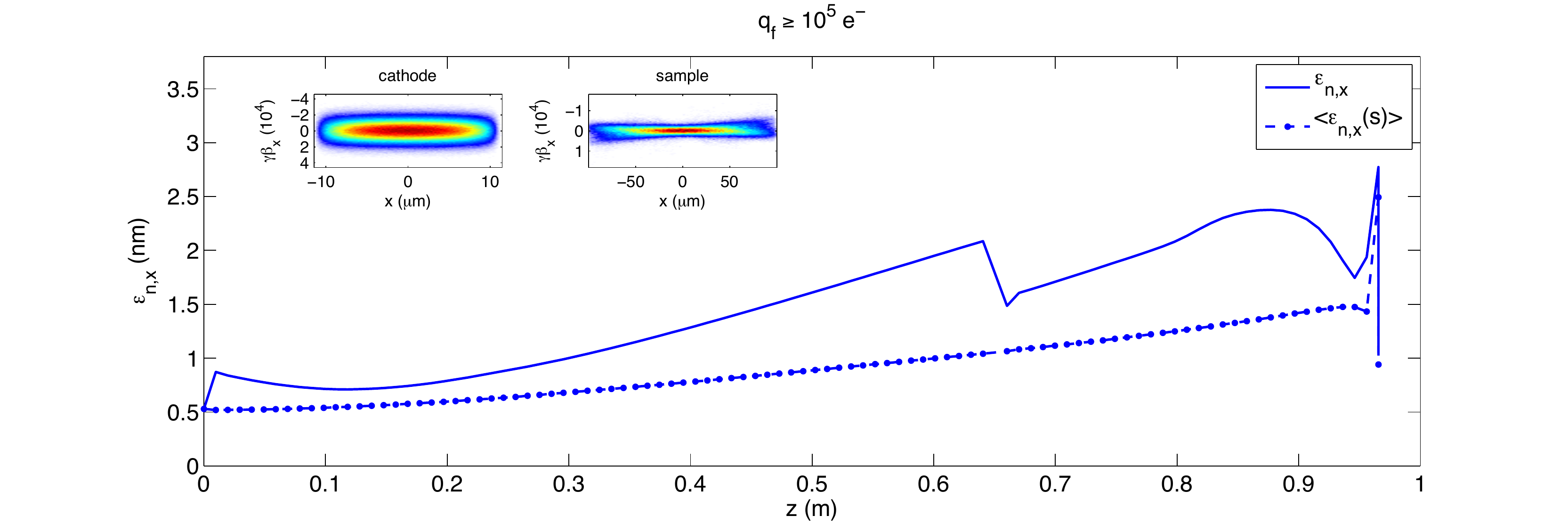}
        }\\ %  ------- End of the first row ----------------------%
        \subfigure[\hspace{0.2cm}]{%
           \label{fig:enx_cgun2}
           \includegraphics[width=0.95\textwidth]{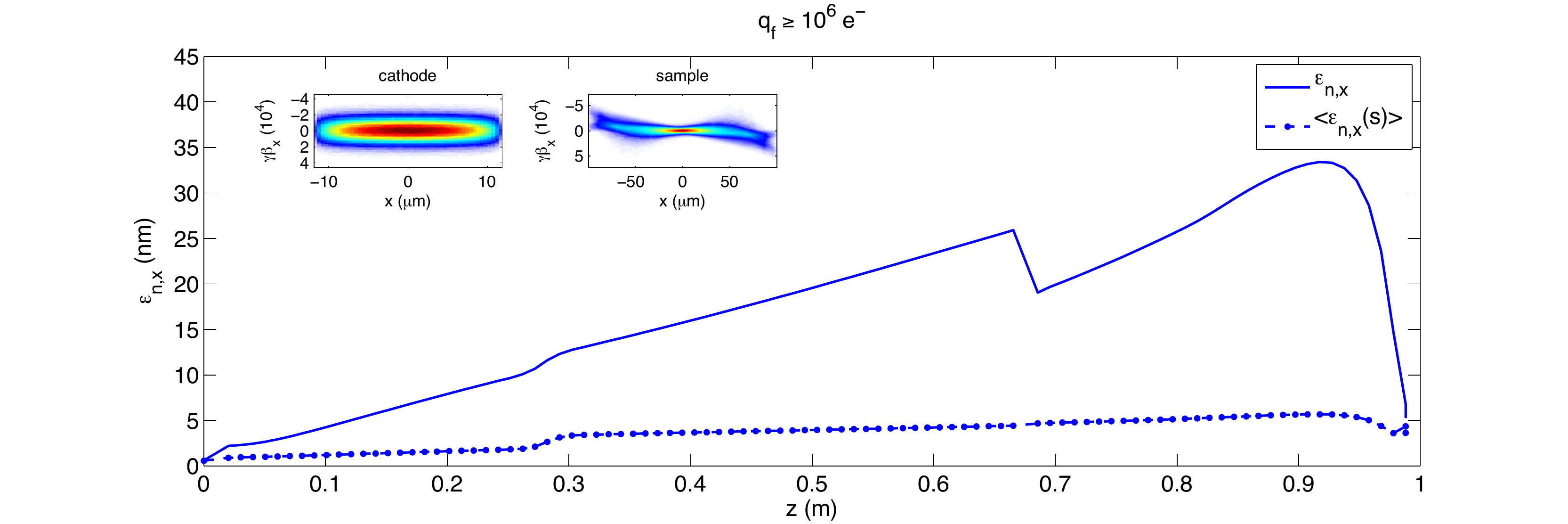}
        }\\ %  ------- End of the first row ----------------------%
    \end{center}
    \caption{%
    \label{fig:enx_1e6}
    Transverse rms projected and average slice emittance along the cyrogun beamline for a final charge of $10^5$ electrons (a) and $10^6$ (b) electrons.  Insets show the transverse phase space distributions at the cathode and sample locations.}%
\end{figure*}

As the emittance performance largely determines the optimal coherence length, we also plot both the transverse projected and average slice emittance along each beamline for each final charge.  For the slice emittance calculation, each simulation was run with 30k macroparticles, and binned using 20 longitudinal slices along the bunch.  The emittance in slice, $\epsilon_{n,x}(s)$, was computed and then the average over the slices taken to get a single number representative of the slice data.  Fig.~\ref{fig:enx_cgun1} shows the emittances computed using the lower final charge for cryogun.
%\begin{figure*}[ht!]
%    \begin{center}
%        \subfigure[\hspace{0.2cm}]{%
%           \label{fig:enx_cgun1}
%           \includegraphics[width=0.95\textwidth]{figs/cgun_enx_1e5}
%        }\\ %  ------- End of the first row ----------------------%
%        \subfigure[\hspace{0.2cm}]{%
%           \label{fig:enx_ncrf1}
%           \includegraphics[width=0.95\textwidth]{figs/ncrf_enx_1e5}
%        }\\ %  ------- End of the first row ----------------------%
%    \end{center}
%    \caption{%
%    \label{fig:enx_1e5}
    %Transverse rms projected and average slice emittance along the cyrogun (a) and NCRF beamlines (b) for a final charge of $10^5$ electrons.  Insets show the transverse phase space distributions at the cathode and sample locations.}%
%\end{figure*}
Shown in the insets are the initial and final horizontal phase spaces in both cases.  The space charge induced rotation of the slices grows the projected emittance along the beamline up to the point of the last solenoid before the sample, which is used to send the beam through a waist, aligning the slices in the process.  We point out that the emittance drops following the eventual slice realignment due to the second solenoid.  However, the emittance blows up again as the beam is compressed longitudinally before being clipped at the sample, after which the emittance is on the order of 1 nm.  %A similar effect from the clipping is seen in RF gun case shown in Fig.~\ref{fig:enx_ncrf1}, where a significant portion of the beam is clipped.  
Fig.~\ref{fig:enx_cgun2} shows the corresponding data for the final charge of $q_f\geq10^6$ electrons.  The dynamics is similar to the lower charge, though emittance is significantly larger along the beamline.  The curvature of the final phase in this case indicates the bunch has experienced non-linear fields along transport, which is verified by the increase of the average slice emittance along the beamline.  %The situation is the opposite for the rf gun set-up, where the emittance is actually smaller in this example, due to the smaller initial charge used (larger transmission fraction).  
Table-\ref{tab:resparams} collects all of the relevant emittance data from these simulations, including the estimate of the coherence using Eqn.~(\ref{eqn:lcxemit}).

\begin{figure}[ht!]
\centering
\includegraphics[width=90mm]{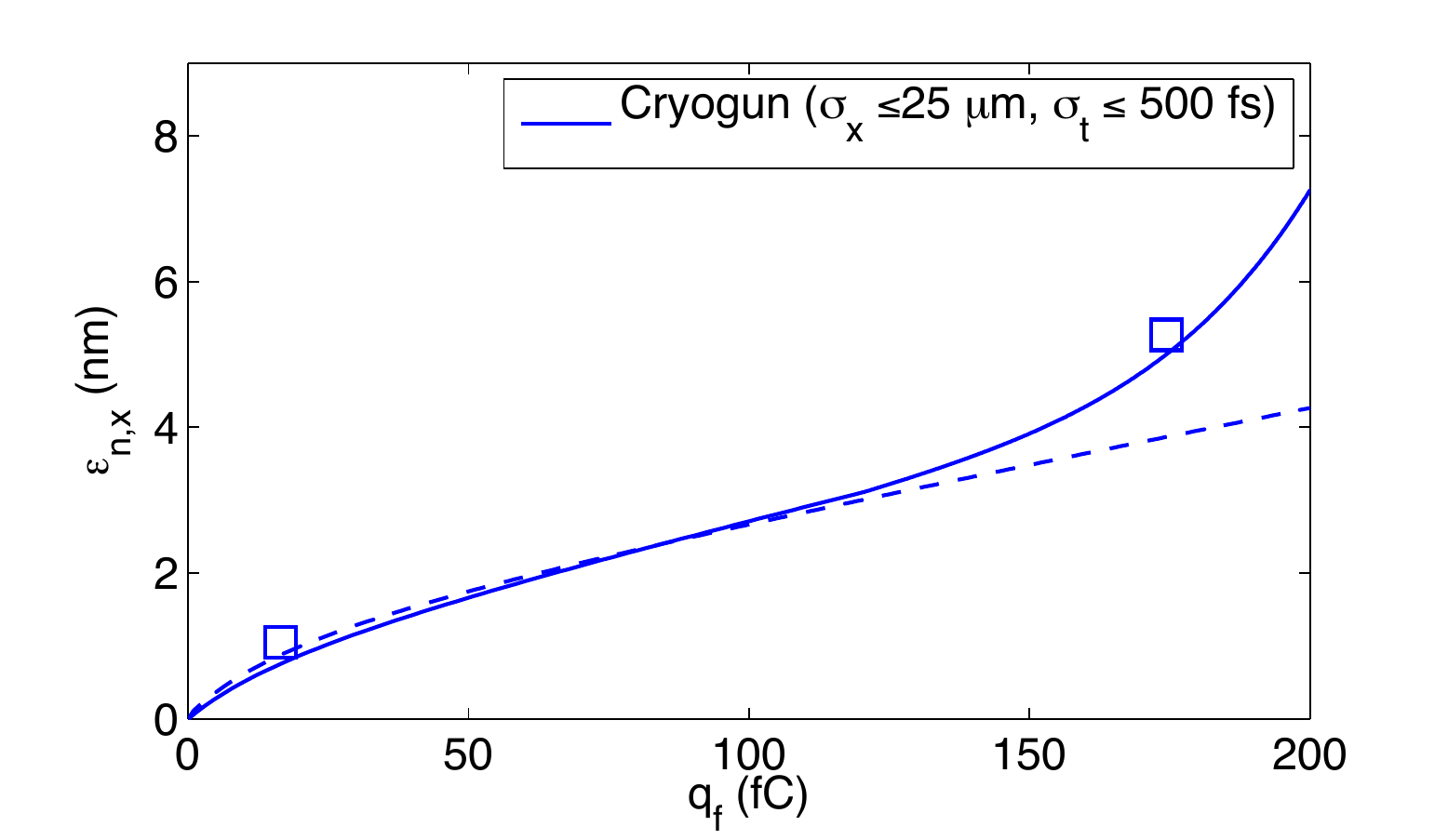}
\caption{Comparison of the optimized emittance with $\sigma_t\leq 500$ fs and the emittance resulting from the coherence length optimizations with $\sigma_t\leq$ 100 (200) fs at $10^5$ ($10^6$) electrons respectively.}
\label{fig:enxfinal}
\end{figure}

Finally, to put the results of the above examples into perspective, we compare the emittances in these results to the optimized emittances for longer bunch lengths. Fig.~\ref{fig:enxfinal} shows the comparison.  As anticipated, the emittance at initial final charges agrees nicely, suggesting the optimizer compensates the requirement of additional bunch compression by clipping out particles (hence the smaller particle transmission at the sample).
%For the larger of the two final charges, the cryogun produces emittances closer to the optimized emittance curves than the RF gun, likely due to use of the separate buncher cavity in the Cryogun beamline.  
These results show that, even when including particle clipping at the sample, the optimized emittances for given final rms transverse and longitudinal beamsizes correctly estimate the coherence length performance.

In this work, we have presented optimized layouts and element settings found using MOGA optimizations of space charge simulations of  a 225 kV DC gun featuring a cryo-cooled photocathode, separate bunching cavity, and two solenoids.  In addition to computing the optimal emittances in each set-up, realistic optimizations of the coherence length as a function final bunch length at the sample, for three sample radii, and allowing for charge clipping at the sample, produced coherence lengths that may be suitable for single-shot UED experiments with final electron charges of $10^6$ electrons.  These results for the optimized coherence length show a significant difference in the emittance and coherence performance when increasing the charge required at the same from $10^5$ to $10^6$ electrons.  In particular, estimates of the scaling of the coherence length fronts suggest the coherence length scales as $q^{-1/2}$ for the largest two sample pinhole radii.  In addition to producing coherence data, these simulations also provide optimized beamline element positions.  Example solutions from the optimum coherence length fronts demonstrate reasonable beam dynamics for $10^5$ and $10^6$ electrons.  Analysis of the optimized coherence lengths shows agreement with the simple formula for the coherence length evaluated at a waist $L_{c,x}\approx\lambdabar_e \sigma_x/\epsilon_{n,x}$.  Estimates of the coherence length using the optimized emittance agree well with the coherence length determined from optimization.  %In conclusion, this work shows that comparable coherence length performance of a cryo-cooled DC gun based beamline with that produced from  a S-band rf gun set-up.

\begin{acknowledgments}
This grant was supported by the NSF, Award PHY 1416318.   
\end{acknowledgments}

\end{document}